\documentclass[iop]{emulateapj}

\usepackage{graphicx}
\usepackage{natbib}
\renewcommand{\vec}[1]{\textbf{\em #1}}
\begin{document}
\title{A Simple Model for the Density Profiles of Isolated Dark Matter Halos}

\author{Eli Visbal$^{1,2}$, Abraham Loeb$^1$, and Lars Hernquist$^1$}
\affiliation{$^1$Institute for Theory $\&$ Computation, Harvard University,
60 Garden Street, Cambridge, MA 02138}
\affiliation{$^2$Jefferson Laboratory of Physics, Harvard University,
Cambridge, MA 02138; evisbal@fas.harvard.edu} 

\begin{abstract}
We explore the possibility that the density profiles of elliptical
galaxies and cold dark matter (CDM) halos found in cosmological
simulations can be understood in terms of the same physical process,
collisionless gravitational collapse.  To investigate this, we study a
simplified model, the collapse of a perfectly cold Plummer sphere.
First, we examine an N-body simulation of this model with particles
constrained to purely radial orbits.  This results in a final state
characterized by a profile slightly steeper than $\rho \propto r^{-2}$
at small radii and 
behaving as $\rho \propto r^{-4}$ at large radii, which can be
understood in terms of simple analytic arguments.  Next, we repeat our
simulation without the restriction of radial orbits.  This results in
a shallower inner density profile, like those found in elliptical
galaxies and CDM halos.  We attribute this change to the radial orbit
instability (ROI) and propose a form of the distribution function (DF)
motivated by a physical picture of collapse.  As evidence of the link
between our model and CDM halos, we find that our collapse simulation
has a final state with pseudo-phase-space density which scales roughly
as $\rho/\sigma^3 \propto r^{-1.875}$, like that observed in CDM halos
from cosmological simulations (Navarro et al. 2010).  The velocity
anisotropy profile is also qualitatively similar to that found near
the centers of these halos.  We argue that the discrepancy at large
radii (where CDM halos scale as $\rho \propto r^{-3}$) is due to the
presence of the cosmological background or continued infall.  This
leads us to predict that the outer CDM halo density profile is not
``universal,'' but instead depends on cosmological environment (be it
an underdense void or overdense region). 

\end{abstract}

\keywords{Cosmology: theory --- dark matter --- Galaxies: elliptical
and lenticular, cD }

\maketitle

\section{Introduction}
Over roughly the past 15 years, considerable attention has been
directed at the possibility that halos in cold dark matter (CDM)
universes may have so-called ``universal'' density profiles.  This
idea was motivated by studies of cosmological N-body simulations which
found that the spherically averaged density profiles of CDM halos
could be accurately fit by the same simple function for a very large
range in mass \citep{1996ApJ...462..563N}.  This discovery led to the
popular NFW profile parameterization,
\begin{equation}
\rho_{\rm NFW}(r) \propto \frac{1}{\left (r/r_s \right ) \left (1+r/r_s \right)^2},
\end{equation}
where $r_s$ is a characteristic length scale.  There have been
numerous attempts to understand the physical nature of the NFW profile
and other similar parameterizations.  The work has mostly focused on
an origin within the cosmological context, investigating the role of
large-scale structure and cosmological infall and accretion
\citep[e.g.][]{2011ApJ...734..100L,2010arXiv1010.2539D,2000ApJ...538..528S,2010PhRvD..82j4044Z,2010PhRvD..82j4045Z}.
These efforts have generally failed to produce a complete physical
description of the processes which lead to the halo density profiles
found in simulations.

It is well known that elliptical galaxies have surface density
distributions described by the de Vaucouleurs $R^{1/4}$ law (or its
generalization the Sersic law)
\citep[e.g.][]{2004AJ....127.1917T,2003AJ....125.2936G}.  A convenient
analytical approximation to the deprojected $R^{1/4}$ law is given by
the Hernquist model \citep{1990ApJ...356..359H}
\begin{equation}
\rho_{\rm H}(r) \propto \frac{1}{\left (r/r_s \right ) \left (1+r/r_s \right)^3}.
\end{equation}
The NFW and Hernquist profiles are strikingly similar, only differing
asymptotically at large radii.  This suggests a link between the
density profiles of elliptical galaxies and CDM halos.  In fact, they
can be accurately fit in projection by the Sersic law or in three
dimensions by a related profile, the
Einasto model, which has been shown to fit CDM halos more accurately
than the NFW parameterization \citep{2005ApJ...624L..85M,2010MNRAS.402...21N}.  This
similarity motivates the hypothesis that the same physical processes
are responsible for the form of the density profiles in both CDM halos
and elliptical galaxies.

This leads us to propose that the formation of CDM halos can be
understood, at least in part, in terms of the physics of
dissipationless gravitational collapse common to elliptical galaxies
and CDM halos.  We can apply this physics to better understand the
density profiles observed in cosmological simulations.  For example,
the analytic arguments made 25 years ago by
\citet{1987IAUS..127..511J} and \citet{1987IAUS..127..339W} imply that
in the outer regions of isolated elliptical galaxies the density
scales as $\rho \propto r^{-4}$.  We suggest, that for dark matter
halos, any deviation from this is due to continued accretion or
because at large radii the cosmological background density dominates.
Thus, to the extent that the structure of halos depends on their
instantaneous accretion state or the local background density, the
notion that CDM halos have ``universal" density profiles is not
accurate.  In a $\Lambda$CDM universe, an isolated halo in a
very low density environment will stop accreting relatively early, and
hence will achieve the limiting form $\rho \propto r^{-4}$.  This is
consistent with the results of the cosmological simulations of
\citet{1991ApJ...378..496D}, where vacuum boundary conditions led to
an outer profile with $\rho \propto r^{-4}$.  We also note that the
recent work of \citet{2011ApJ...734..100L} and
\citet{2011MNRAS.414.3044V} demonstrates that for scale-free models,
the inner halo profile is sensitive to the shape of the collapsing
perturbation.  This suggests that the inner profile of halos could be
sensitive to environment as well.

In this paper, the first in a series of two, we explore the physics of
dissipationless gravitational collapse and how it applies to the
formation of CDM halos.  We accomplish this with numerical experiments
and analytic arguments based on a simple toy-model, the collapse of a
perfectly cold (zero initial kinetic energy) Plummer sphere.  In Paper
2 (in preparation) we will study the impact of cosmological effects,
such as continued accretion.  This will include analyzing cosmological
N-body simulations to study the expected systematic dependence of the
detailed profile on cosmological environment, be it an underdense void
or an overdense region. 

We begin the study of our collapsing Plummer sphere by performing a
simulation of purely radial collapse in which non-radial motions are
suppressed.  This results in a final equilibrium state with a density
profile slightly steeper than $\rho \propto r^{-2}$ near the center
and $\rho \propto r^{-4}$ at large radii.  However, a system composed
entirely of particles on purely radial orbits will be dynamically
unstable once non-radial motions are allowed, owing to the radial
orbit instability (ROI), an effect first noted by
\citet{1986ApJ...300..112B} and \citet{1985MNRAS.217..787M}.  When we
remove the restriction of purely radial orbits we find a final state
with a shallower inner profile more similar to that found in
elliptical galaxies and CDM halos.  The possibility that the ROI could
play a role in structuring CDM halos has been emphasized already by
\citet{2008ApJ...685..739B}.  Below, we examine this idea in greater
depth by postulating a physically motivated form for the distribution
function (DF) of systems which have relaxed through the ROI.  As
evidence of the link between our simple collapse model and CDM halos,
we find that the final state has a pseudo-phase-space density
$\rho(r)/\sigma(r)^3 \propto r^{-1.875}$, as observed in halos in
N-body cosmological simulations \citep{2010MNRAS.402...21N} and
derived in self-similar secondary infall models
\citep{1985ApJS...58...39B}.

This paper is structured as follows.  In Section 2 we present the
simulation of our toy-model of gravitational collapse with particles
restricted to purely radial orbits.  In the following section, we
discuss the outer density profile of this numerical experiment and how
it relates to CDM halos and elliptical galaxies.  We consider the
inner density profile in Section 4.  Here we re-simulate the collapse
from Section 2 without the restriction of purely radial orbits, and
discuss the impact of the ROI.  In Section 5 we propose a physically
motivated DF for the final state of systems
formed through initially radial gravitational collapse.  We discuss
our results in the context of previous work in Section 6 and give our
main conclusions in Section 7.

Throughout the paper, the data from our
numerical experiments are presented in the internal units we use in
the GADGET N-body code \citep{2005MNRAS.364.1105S}.  Distances are
given in units of $1 {\rm kpc}$, velocities in units of $1 {\rm
km~s^{-1}}$, and mass in units of $10^{10} M_\odot$.
Other units (e.g. time, energy, density) are given by the relevant
combination of these three quantities.

\section{A Simple Example} 
\begin{figure*}
\centering
\includegraphics[width=3in]{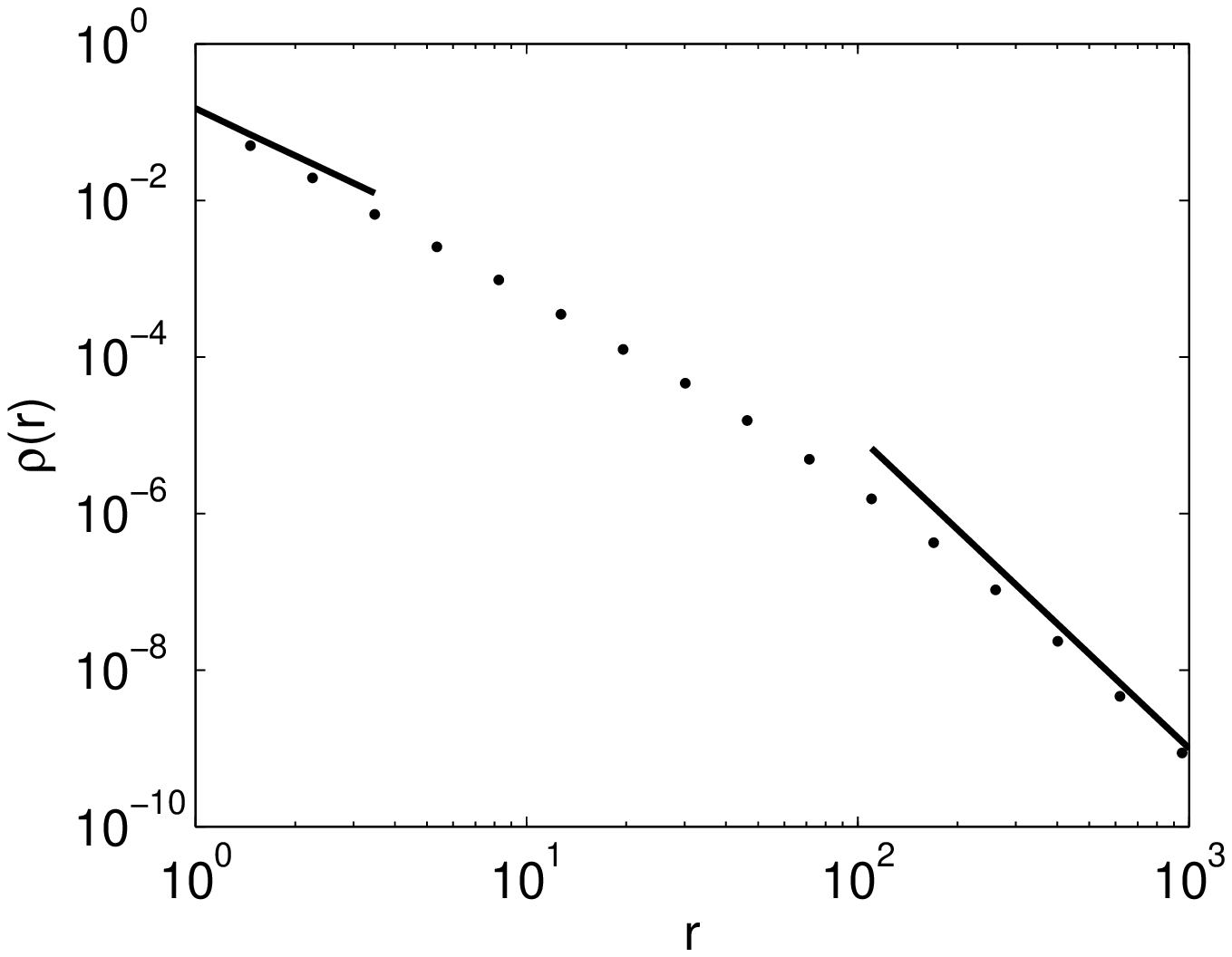}
\includegraphics[width=3in]{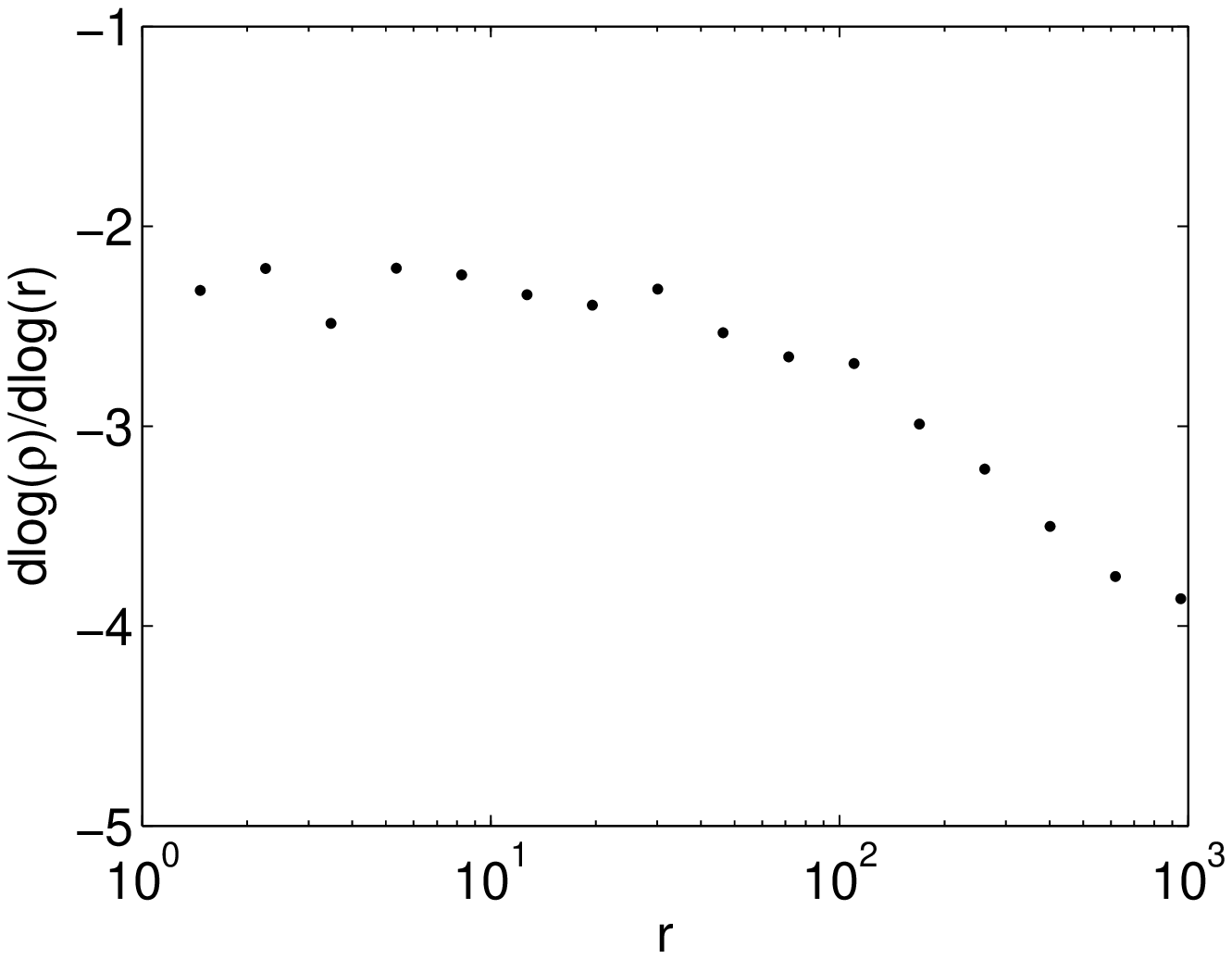}
\caption{\label{fig1r} The density profile (left panel) and
logarithmic slope (right panel) of the final state from our simulation
of purely radial collapse starting from a perfectly cold Plummer
sphere (see Section 2 for more details).  The solid lines indicate $\rho
\propto r^{-2}$ at small radii and $\rho \propto r^{-4}$ at large radii.}
\end{figure*}

We begin the exploration of our simplified model with an N-body
simulation of gravitational collapse where particle orbits are
constrained to be completely radial.  The simulation was initialized
as a perfectly cold (all particles with zero velocity) Plummer sphere
with scale length of 100 and total mass of 100 in the code units
described above.  These choices are unimportant, as neither the total
mass nor the scale length affect the shape of the density profile for
the final state.  The system was evolved in time with particles only
subjected to the component of the gravitational force in the radial
direction.  This was accomplished by modifying the publicly available
N-body code GADGET \citep{2005MNRAS.364.1105S}.  The simulation was
run until it reached an equilibrium state, with the density profile
plotted in Figure \ref{fig1r}.  We have achieved convergence in the
sense that the final state is not sensitive to the number of particles
included in the simulation.  For the results shown, we used $10^5$
particles which were initially distributed by randomly drawing from a
probability density function (PDF) proportional to the Plummer
profile.  A gravitational softening length of $\epsilon_g = 0.3$ was
adopted; reducing this value did not affect the results on the scales
of interest.  Note that the specific choice of a Plummer sphere does
not have a strong impact on the final state.  We find similar outcomes
with other initial density profiles (e.g. Gaussian or Hernquist).

Our simulation results in an equilibrium state slightly steeper than
$\rho \propto r^{-2}$ at small radii and $\rho \propto r^{-4}$ at
large radii.  We can show that in general, a density profile formed
through purely radial collapse cannot be shallower than $\rho \propto
r^{-2}$ using fairly simple analytical arguments
\citep{2010arXiv1010.2539D}.  In the central regions, the time
averaged contribution to the total density profile from each particle
is
\begin{equation}
\label{rad_arg}
\rho_p(r) \propto \frac{dt(r)}{4\pi r^2 dr}  \propto \frac{1}{r^2} \frac{dt}{dr},
\end{equation}
where $dt(r)$ is the time the particle spends between radii $r$ and
$r+dr$.  For a density profile shallower than $\rho \propto r^{-2}$,
$dt/dr$ is a constant for sufficiently small $r$.  It follows
from equation (\ref{rad_arg}) that central density profiles must always
be at least as steep as $\rho \propto r^{-2}$ in the purely radial
case.


The large radius limit can be explained with the following argument
\citep{1987IAUS..127..339W,1987IAUS..127..511J}.  In the final state
of a system formed through gravitational collapse, the outer envelope
will be populated by particles which have been scattered into weakly
bound orbits.  The size of these orbits will be inversely proportional
to the energy, $E$.  The number of particles between $E$ and $E+dE$,
$N(E)dE$, is expected to be non-zero and continuous near $E=0$.  For
the density, this implies
\begin{equation}
\rho(r) = \frac{N(E)}{4\pi r^2}\frac{dE}{dr} \propto \frac{N(GM/r)}{r^2} \frac{d(GM/r)}{dr} \propto r^{-4}.
\end{equation} 
Note that this result was derived in the context of elliptical
galaxies and has been completely ignored in the literature on
cosmological CDM halo formation.

Although the above discussion is (intentionally) highly idealized, it
demonstrates some important consequences.  First, there is nothing
mysterious or pathological about density profiles which diverge as $r
\rightarrow 0$.  Second, the large-$r$ behavior is consistent with the
analytical arguments of \citet{1987IAUS..127..339W} and
\citet{1987IAUS..127..511J}.  Note that while ``violent-relaxation"
\citep{1967MNRAS.136..101L} plays a role in the evolution by
redistributing energy among the particles, and hence promoting a small
fraction to unbound orbits, it does not in detail determine the
limiting behavior at small and large radii, and hence does not justify
efforts to understand the nature of the generic density profiles based
on thermodynamic arguments \citep[e.g.][]{1992ApJ...397L..75S}.
Finally, the lack of sensitivity to the number of particles indicates
that the outcome has been determined by the physics of collisionless
dynamics.  We argue that the physical processes acting in this example
are, at least in part, responsible for the origin of the density
profiles of dark matter halos.  While the density profile obtained
from this example does not match the NFW form, we can nevertheless use
it as a starting point for the more detailed study described in the
following sections.

Before moving on, we note that the final state in this example can be
approximated by the density profile of the Jaffe model,
\begin{equation}
\rho_{\rm J}(r) \propto \frac{1}{\left(r/r_s \right)^2 \left (1+ r/r_s \right)^2}.
\end{equation}
However unlike the isotropic Jaffe model, the velocities are purely
radial and therefore are dynamically unstable when particles are
permitted to have non-radial velocities (as we explore below).  Note
that the DF for a Jaffe model density profile
constrained to purely radial orbits can be expressed analytically.  We
derive this expression in the Appendix.

\section{The Outer Density Profile}
According to the arguments put forward by \citet{1987IAUS..127..339W}
and \citet{1987IAUS..127..511J}, an isolated collisionless system
evolving into equilibrium through collapse should eventually achieve
the limiting form $\rho \propto r^{-4}$ at large radii.  This expected
outcome is in agreement with the simple numerical experiment presented
in Section 2.  Note that the analytic argument is insensitive to orbital
anisotropy, and therefore is unaffected by the imposed constraint of
purely radial orbits.  We point out that this large radius behavior
should also occur in dark matter halos once cosmological infall
terminates. \citet{1991ApJ...378..496D} originally demonstrated this
using cosmological simulations in which they employed vacuum boundary
conditions around individual objects which naturally led to a
cessation of late time accretion.

We suggest that the NFW profile at large radii, $\rho \propto r^{-3}$,
does not directly reflect the physical processes that form CDM halos.
We find in our collapse simulations (both the purely radial case
discussed above and the generalized case discussed below) that the
density profile reaches $\rho \propto r^{-4}$ roughly two orders of
magnitude beyond the radius where $d{\rm ln}(\rho)/d{\rm ln}(r) \approx -2$.
For most CDM halos in the cosmological context, the background density
will dominate at such large radii.  Thus, CDM halos may not extend far
enough before falling below the background density to achieve $\rho
\propto r^{-4}$.  Additionally, continued infall in high density
regions could alter the outer profile.  This finding has important
implications for understanding the formation of halos in a
cosmological context.  Low-density regions will behave as open
universes, implying that accretion there will halt earlier than in
higher density environments.  Some regions may be underdense enough to
permit halos to reach the asymptotic form $\rho \propto r^{-4}$.
Therefore, it is expected that the outer profiles of halos may exhibit
a variety of behaviors, depending on environment and redshift, so in
detail the outer profiles of dark matter halos will not be
characterizing by a universal form.  This will be explored in detail
in paper 2 (in preparation).

\section{The Inner Density Profile and the Radial Orbit Instability}
\begin{figure*}
\centering
\includegraphics[width=3in]{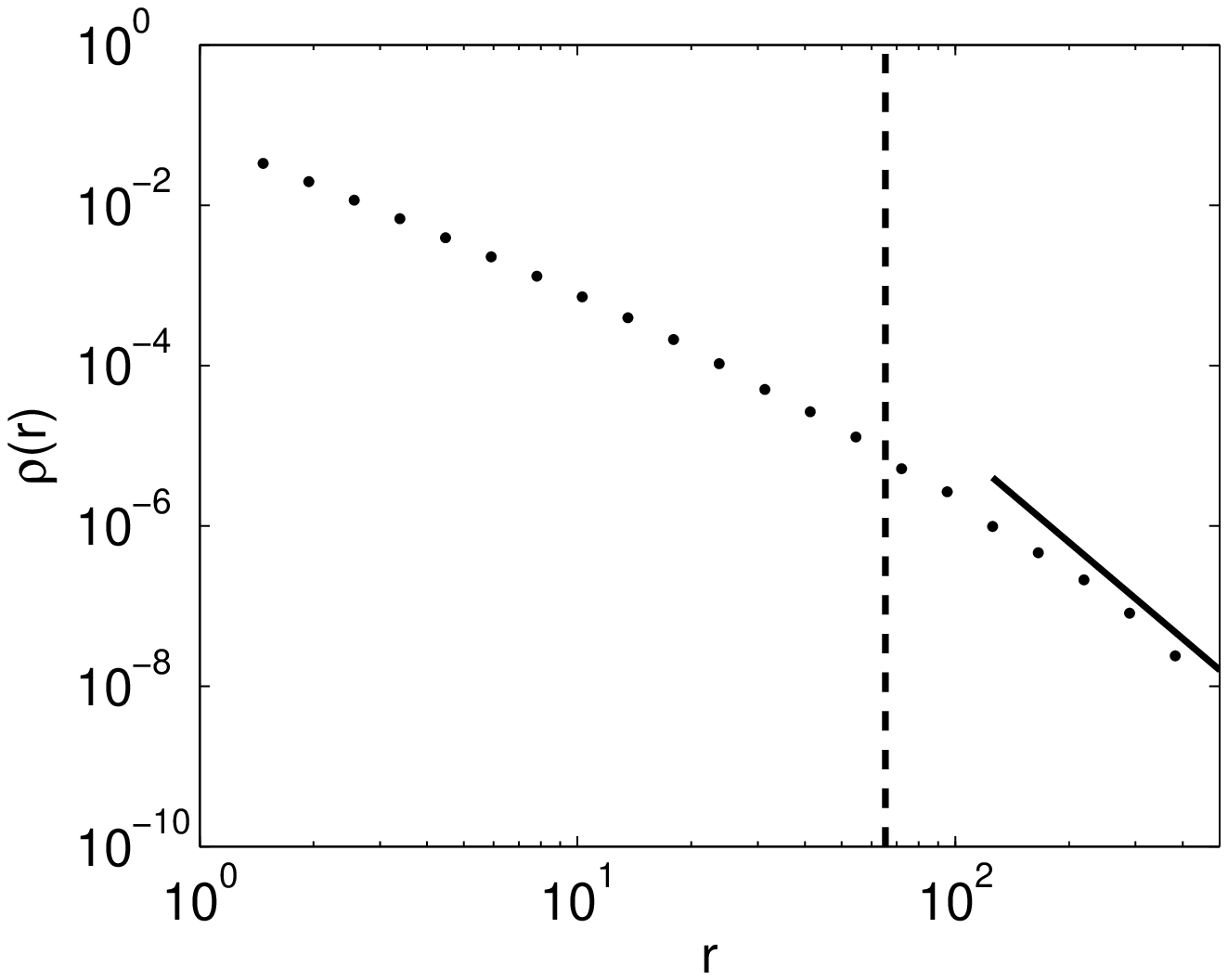}
\includegraphics[width=3in]{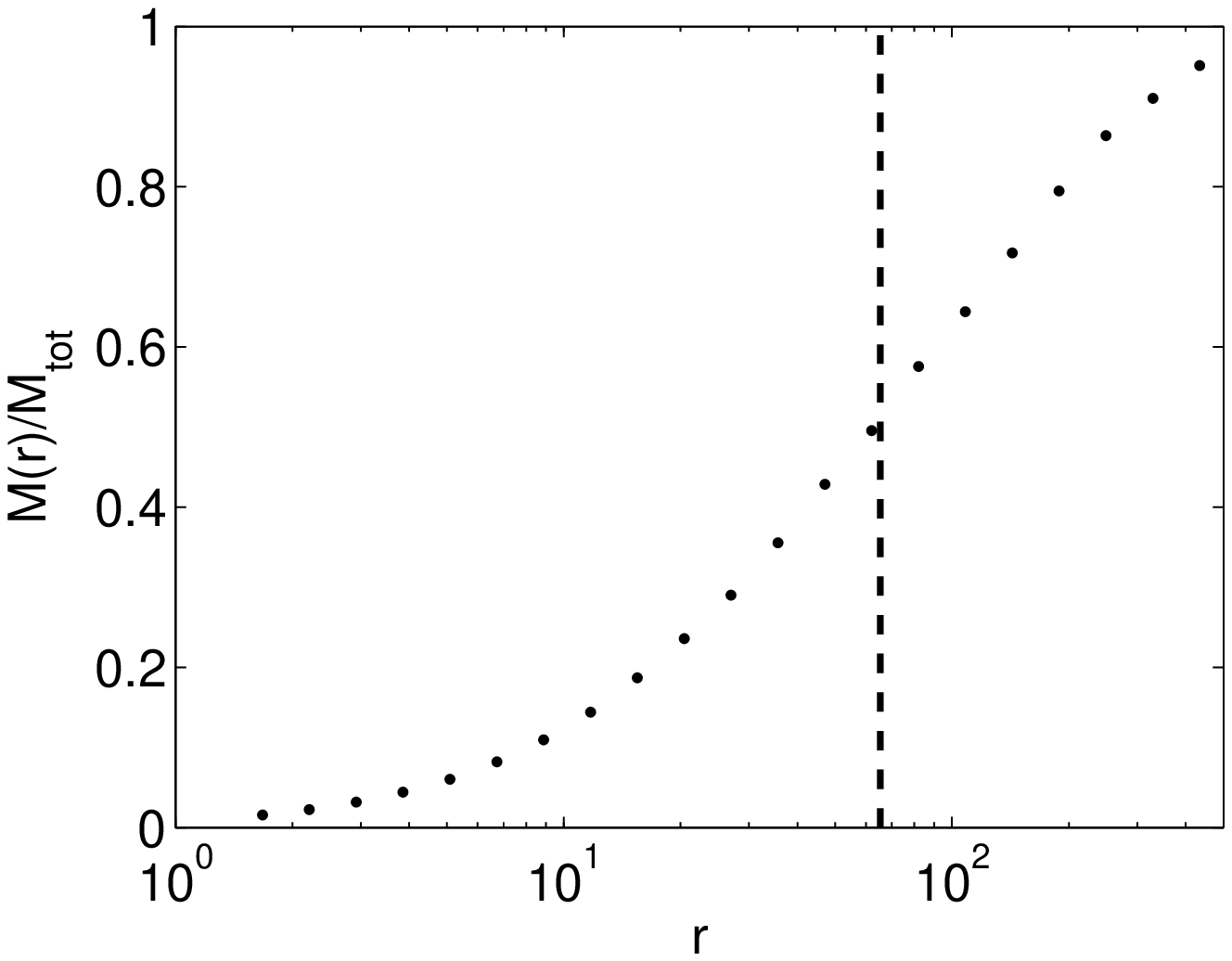}
\includegraphics[width=3in]{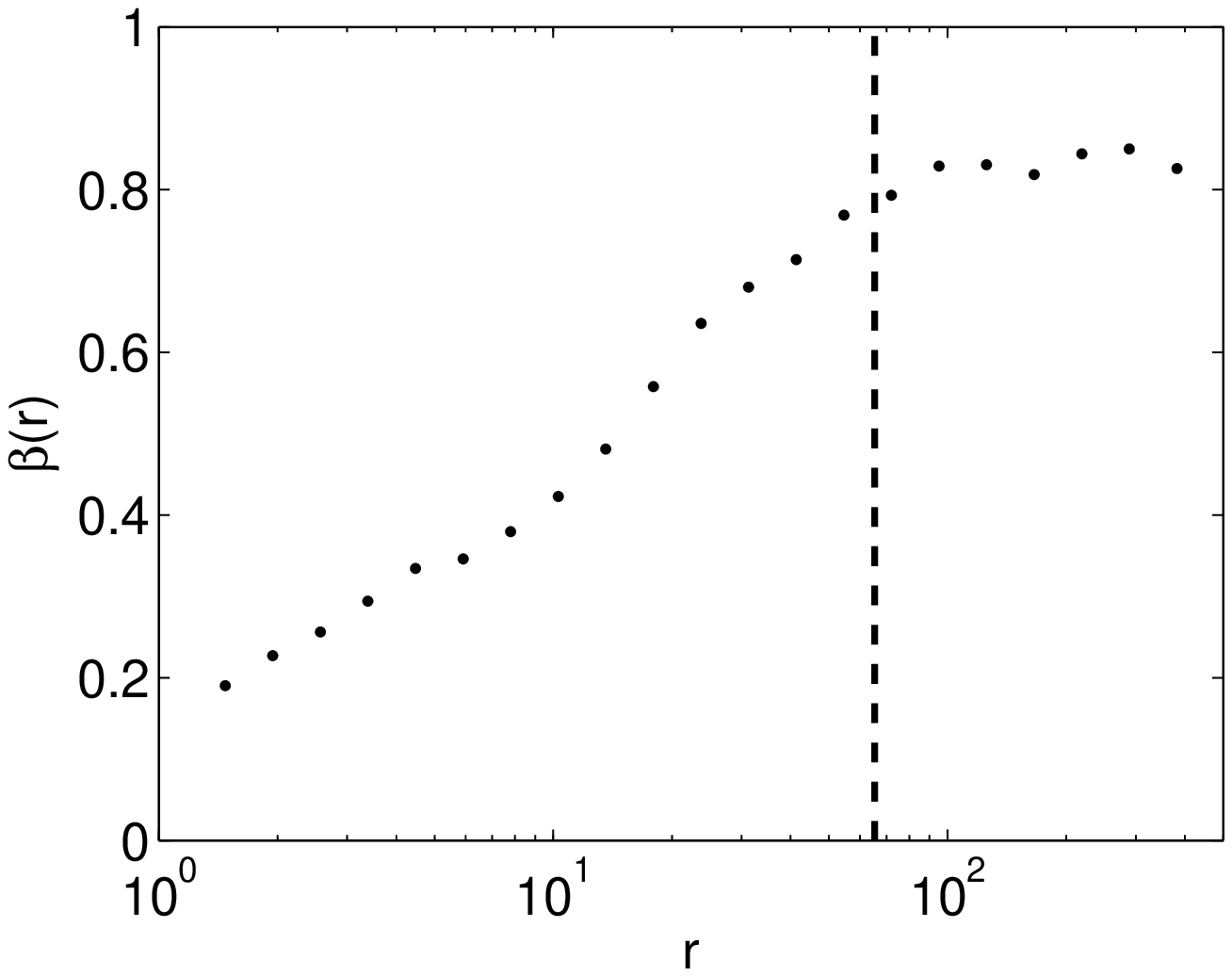}
\includegraphics[width=3in]{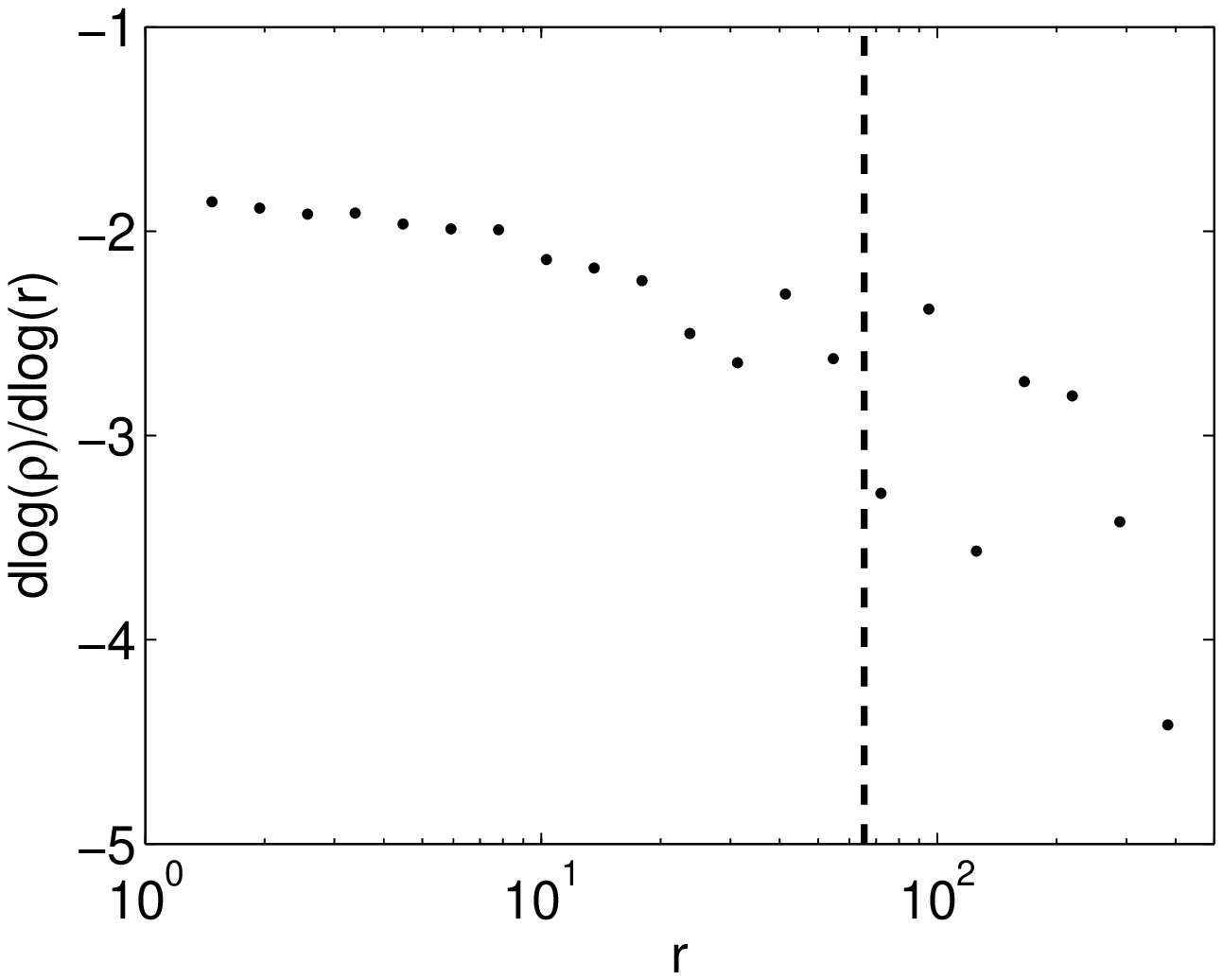}
\caption{\label{fig3r} The final state of the N-body simulation of
collapse from a cold (zero velocities) Plummer sphere described in
Section 4.  We plot radial profiles of the density (upper left),
cumulative mass interior to a radius $r$ (upper right), anisotropy
parameter (lower left), and logarithmic slope (lower right).  The
solid line indicates $\rho \propto r^{-4}$ and the dashed line marks
the half-mass radius.}
\end{figure*}

To investigate the small-radii properties of equilibrium systems
formed through gravitational collapse, we generalize the numerical
experiment described in Section 2.  We perform the same N-body
simulation of radial collapse, but relax the restriction that the
particles move on purely radial orbits.  Here we use $2 \times 10^6$
particles and a softening length of $\epsilon_g = 0.1$.  We find that
we have achieved convergence in the sense that increasing particle
number or decreasing softening length do not impact our results on the
scales of interest.  Since our results are not sensitive to particle
number, we conclude that 2-body relaxation does not impact our results
and the system is effectively collisionless.  As in the purely radial
simulation, different initial density profiles result in similar final
equilibrium states.
The results of this simulation are presented in Figure \ref{fig3r}.
Note that the simulation was run roughly five times the dynamical time
at the maximum radius shown.  While we may not have reached perfect
equilibrium at this radii, we do not expect substantial changes at
longer times.

Initially the system has mainly radial orbits (as in the first
numerical experiment).  However, in response to small non-radial
perturbations, purely radial orbits are dynamically unstable.
The system achieves a different
equilibrium in which the orbits in the inner regions are isotropic,
but those in the outer regions, where the system is not strongly
self-gravitating, are still mainly radial.  Roughly, the transition
occurs at the half-mass radius of the new configuration.  This is
shown in Figure \ref{fig3r}, where we plot the radial profile of 
the anisotropy parameter,
\begin{equation}
\beta(r) = 1 - \frac{\sigma_t(r)^2}{2 \sigma_r(r)^2}, 
\end{equation}
where $\sigma_r(r)^2$ and $\sigma_t(r)^2$ are, respectively, the
radial and the tangential variance in the velocity as a function of
radius ($\beta=0$ corresponds to isotropic and $\beta=1$ to radial).
In the final state, the system has an inner density profile shallower
than $r^{-2}$, but the outer regions are unchanged compared to the
purely radial case, still having $\rho \propto r^{-4}$. The particles
in this outer region (well outside the half-mass radius) respond to
the gravitational potential dictated by the inner region, and their
orbits are therefore stable. The inner core (inside the half mass
radius), on the other hand, is self-gravitating and subject to the
ROI.  The resulting central density cusp is now more similar to those
which characterize CDM halos and the deprojected $R^{1/4}$ law,
supporting the notion that CDM halos and elliptical galaxies
are shaped by similar processes.

\begin{figure}
\centering
\includegraphics[width=3in]{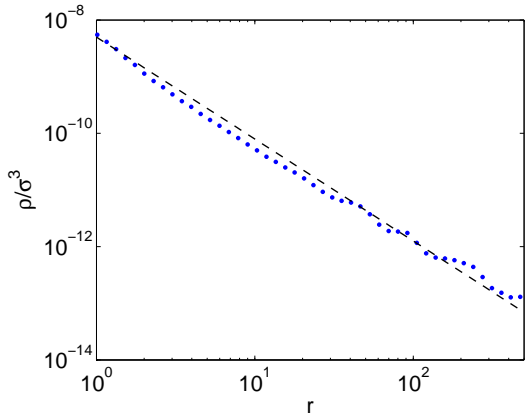}
\caption{\label{fig_phase} The pseudo-phase-space density,
$\rho/\sigma^3$, of the final state of the collapse simulation
described in Section 4 (points).  For reference we plot $\rho/\sigma^3
\propto r^{-1.875}$, which is found in CDM halos from cosmological
N-body simulations. }
\end{figure}

As further evidence of the relationship between the physics of
gravitational collapse in this example and CDM halos we observe that
the pseudo-phase-space density of the final state scales roughly as
$\rho/\sigma^3 \propto r^{-1.875}$, matching that in halos from
cosmological simulations \citep{2010MNRAS.402...21N}.  This is plotted
in Figure \ref{fig_phase}.  We also find that the velocity anisotropy
profile scales roughly as $\beta(r) \propto {\rm ln}(r)$ near the
center, qualitatively similar to that seen in halos from cosmological
N-body simulations.

\section{A Distribution Function for the Relaxed System} 
The numerical experiments and analytic arguments presented in the
preceding sections have led us to propose a simple scenario for how
collisionless systems are formed through gravitational collapse.  The
outer profile is determined by the redistribution of energy among
particles through violent-relaxation, which promotes a small fraction
of the mass onto unbound orbits.  If no further infall occurs, the
density profile will asymptote to $\rho \propto r^{-4}$ as predicted
by \citet{1987IAUS..127..339W} and \cite{1987IAUS..127..511J}.  If
infall has not halted, the system will not be in equilibrium and a
different density profile may be measured at large radii.  For a
nearly radial infall, the ROI will operate causing the system to
evolve such that the inner region is characterized by orbit isotropy
and a density profile shallower than $\rho \propto r^{-2}$ (not
permitted in the purely radial case).  If the collapsing particles
initially have significant non-radial motions, the final profile will
be even shallower.

To understand the physical nature of systems formed through this
scenario, we have performed an analysis of the phase-space
distribution of the final state in the numerical example of Section 4.
Note that because the system is anisotropic, the DF cannot be
expressed as a function of energy alone.  We propose the following
form of the DF (guided by \citet{1991MNRAS.250..812G}),
\begin{equation}
f(E,L) = f(E,x) = g(E)h(x,E),
\end{equation}
where $x=L/L_c(E)$ is the angular momentum divided by the angular
momentum of a particle on a circular orbit with energy $E$.  The
values of $x$ range between $0$ and $1$.  The functions $g$ and $h$ are
given by
\begin{equation}
\label{gfunc}
g(E) = A \exp(-\beta_E E)(-E)^c,
\end{equation}
and
\begin{equation}
\label{hfunc}
h(x,E) = \exp(-x^2/q(E)^2),
\end{equation}
where $A$, $\beta_E$, and $c$ are free parameters.  We expect $q(E)$ to be a
function which increases with $\left | E \right |$ and parameterize it
with a broken power law as explained below.

This model's main features can be explained in terms of the collapse
scenario described above.  The function $h(x,E)$ determines how
anisotropic orbits of a given energy are, with a delta function in $x$
corresponding to purely radial and a constant function corresponding
to isotropic.  We expect that $q(E)$ gets larger for increasing $\left
| E \right |$ such that $h(x,E)$ transitions from a delta function for
the least tightly bound particles to a constant for the most tightly
bound particles.  The most tightly bound particles are those which
make up the inner profile and have been isotropized through the ROI,
while the more weakly bound particles are unaffected by this
instability and are found mostly in the outskirts of the system on
radial orbits.  The energy dependence of the DF in equation
(\ref{gfunc}) is a Maxwellian distribution for large $\left | E \right
|$, but modified by the factor $(-E)^c$ for the least tightly bound
particles.  We can interpret this as a Gaussian distribution
(representing a favorable statistical state), modified at high
energies due to particles being scattered into unbound orbits.  The
values of the parameters in our model can also be related to the bulk
properties of the system.  Our expectation is that the parameter
$1/\beta_e$ is of the order of the kinetic energy per particle.  The
transition in $q(E)$ which changes $h(x,E)$ from a delta function to a
constant should happen roughly at the mean energy of particles found
at the half-mass radius of the system.  This corresponds to the notion
that the ROI isotropizes roughly the inner half of the system and
leaves the outer envelope, which is not strongly self-gravitating,
mostly radial.

\begin{figure*}
\centering
\includegraphics[width=3in]{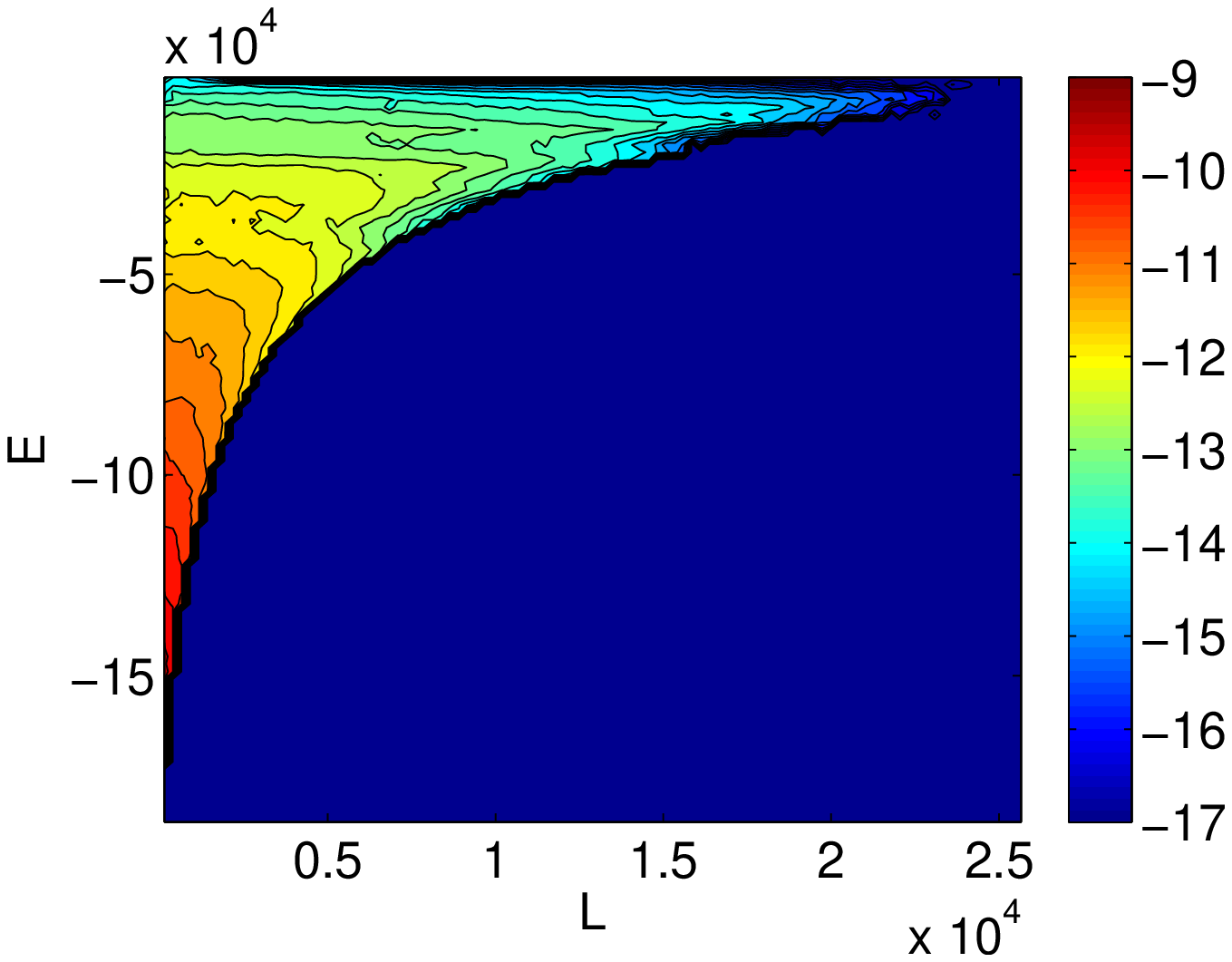}
\includegraphics[width=3in]{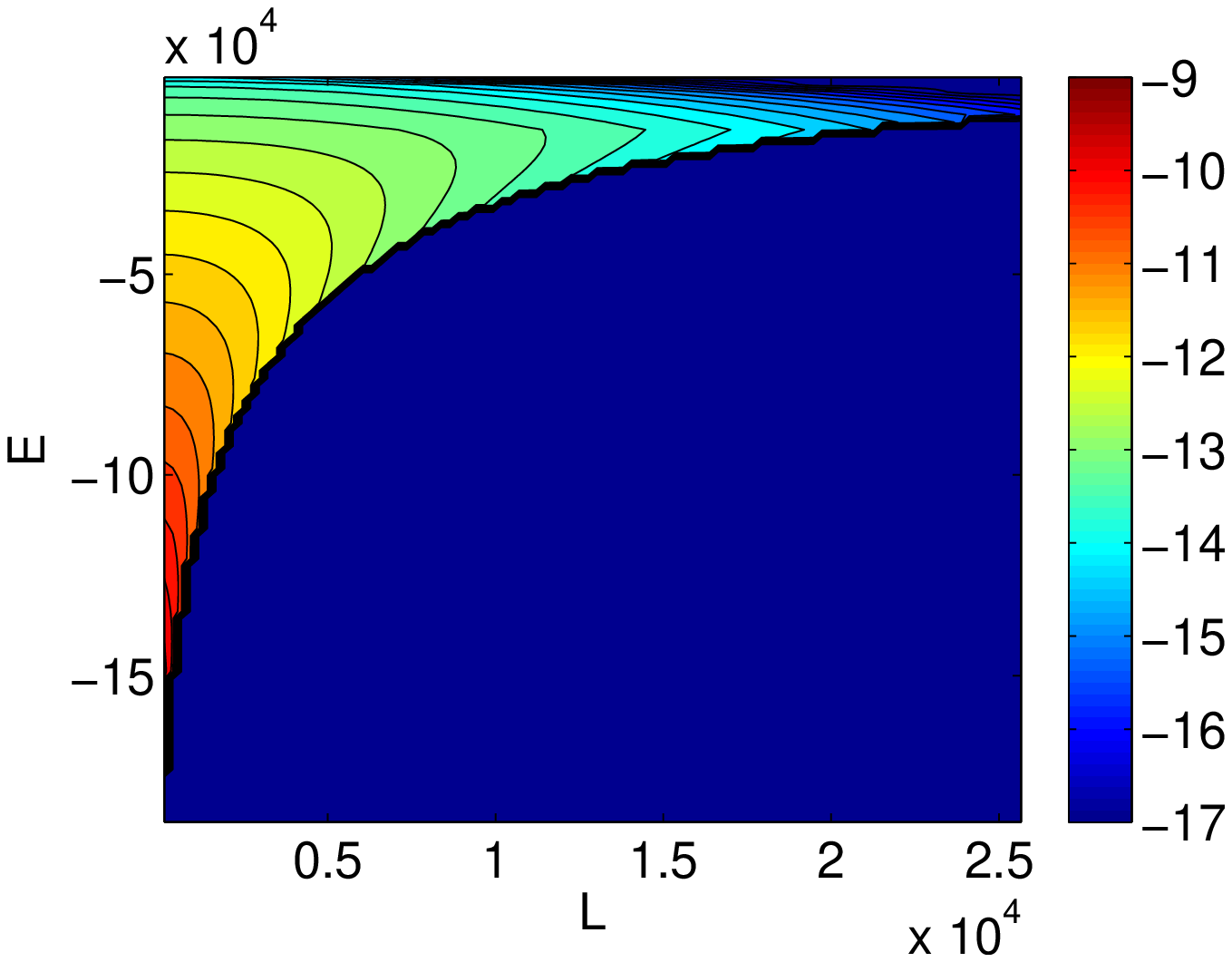}
\includegraphics[width=3in]{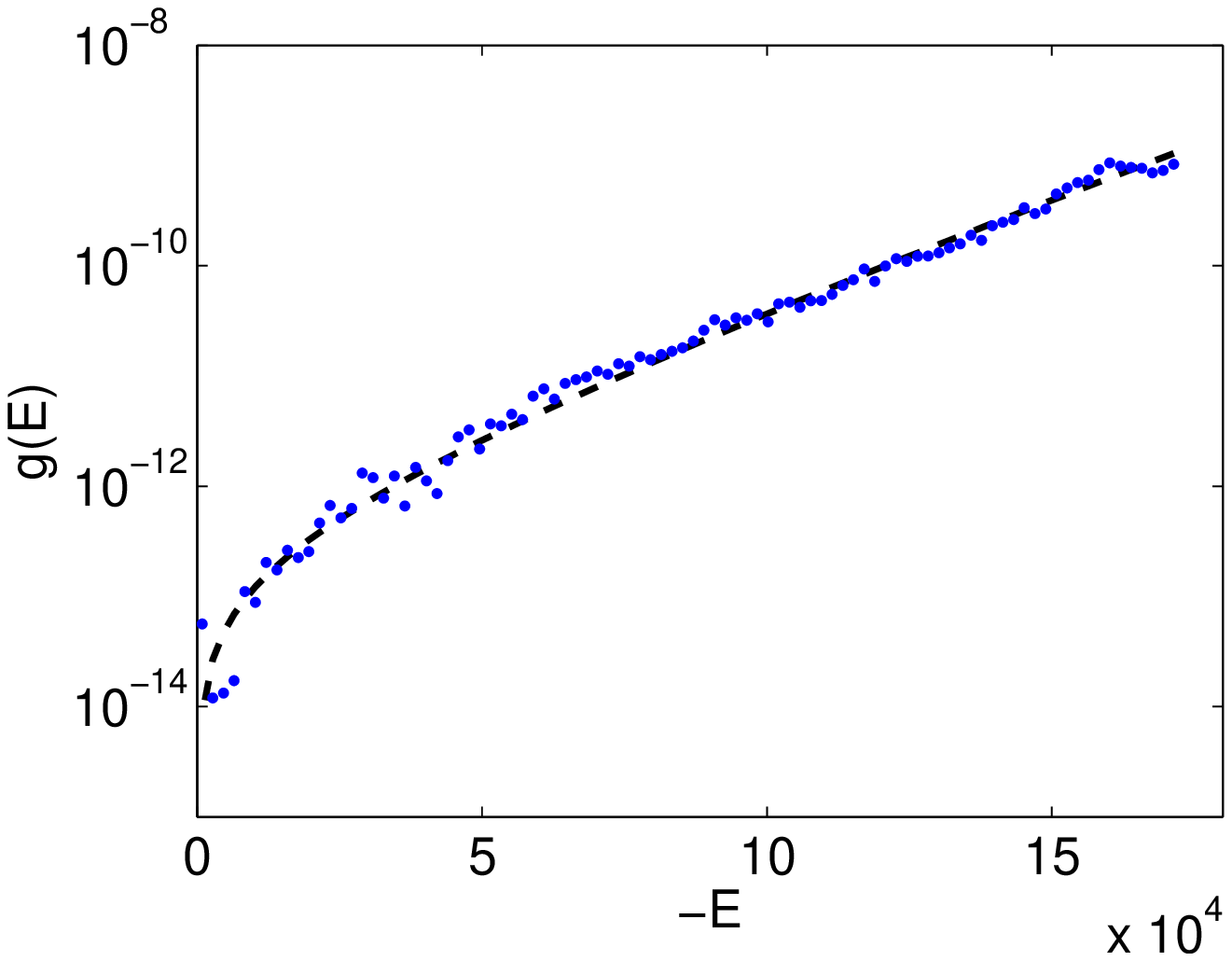}
\includegraphics[width=3in]{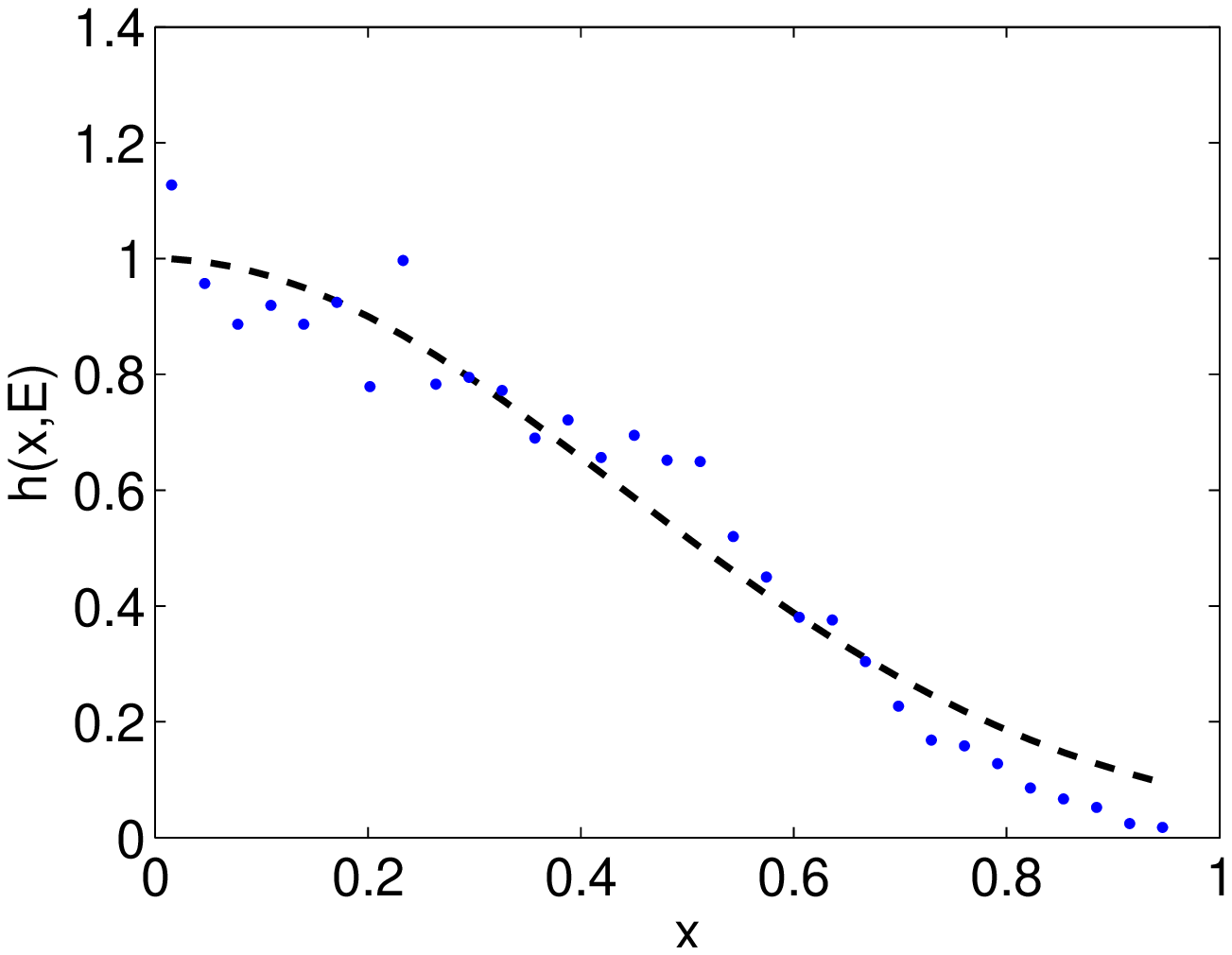}
\caption{\label{fig4r} A comparison of the DF from
the N-body simulation of gravitational collapse described in Section 4
and the model DF described in Section 5.  We plot contours of the DF,
$\log(f(E,L))$, from the simulation (upper left panel) and the parameterized
model (upper right panel).  In the lower left panel we plot $g(E)$,
given by equation (\ref{gfunc}), for the simulation data (blue
points) and the model (dashed line).  In the lower right panel we plot
$h(x,E=-3.8\times 10^4)$, given by equation (\ref{hfunc}).  The values
for the model parameters are listed in the main text.}
\end{figure*}

We compare the DF measured directly from the simulation with a
parameterization of this model.  Values of $c=0.9275$ and $\beta_E = 4
\times 10^{-5}$ are found to match the simulation well (see Figure
\ref{fig4r}).  As expected, the value of $\beta_E$ also corresponds
roughly to the reciprocal of the mean kinetic energy, $\bar{v}^2/2 =
10^{4}$.  
For $q(E)$ we use a broken power law of the form $q(E) = a(-E)^b$.
For $\ln(-E)< 9.55$, $a=0.00012$ and $b=0.87$, while for $9.55< \ln(-E)
< 11.9$, $a=0.07$ and $b=0.21$.  When $\ln(-E)>11.9$, we assume $a=4
\times 10^{-11}$ and $b = 2$.  This was determined by fitting for
$q(E)$ separately in each energy bin and then using the power law as a
simple way to parameterize the values.  This procedure provides a
reasonably good match to the simulation data and has the limiting
behavior we expect for high and low $E$.  We compare contours of the
DF, $f(E,L)$, from the simulation and this parameterization of the
model in Figure \ref{fig4r}. Clearly, there is very good agreement.

\begin{figure*}
\centering
\includegraphics[width=3in]{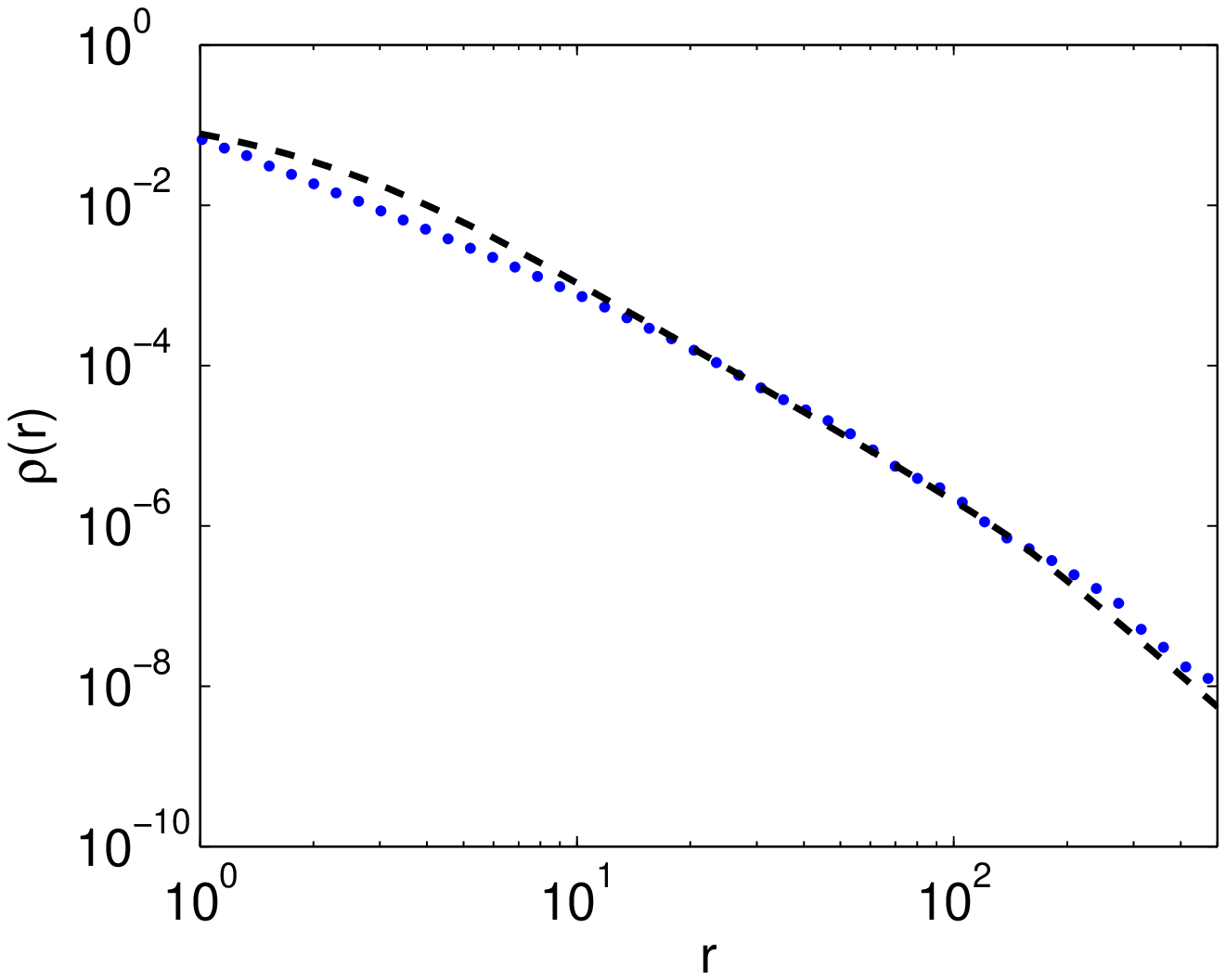}
\includegraphics[width=3in]{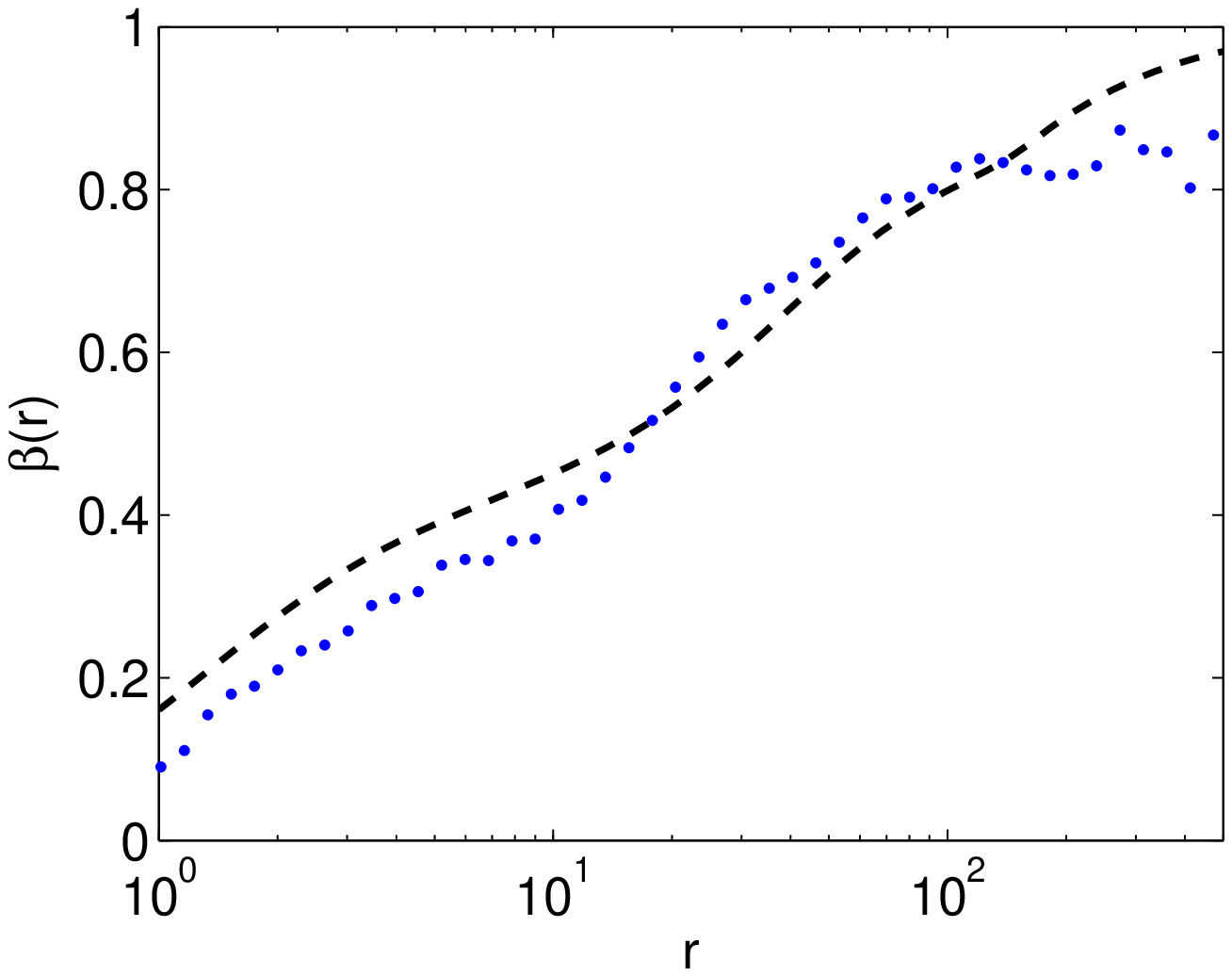}
\caption{\label{fig5r} A comparison between the numerical simulation
described in Section 4 and the model DF described in Section 5.  We plot the
density profile (left panel) and the velocity anisotropy (right panel)
from the simulation (blue points) and the values implied by the model
DF (dashed lines).  The values for the model parameters used are listed
in the main text.}
\end{figure*}

In addition to fitting the phase-space data from the numerical
experiment we also take this DF parameterization and self-consistently
compute the density and anisotropy profiles of the system.  This was
accomplished iteratively with the following expression:
\begin{equation}
\rho_{n+1}(r) = \int d^3\vec{v} f(E=v^2/2+\Phi(r),L=v_t r),
\end{equation}
where the gravitational potential $\Phi(r)$ is computed using the
density from the previous step, $\rho_n$.  At each step the
normalization of the DF is set to fix the innermost
value of $\rho(r)$ to a constant value.  We use the density profile
from the simulation as $\rho(r)_1$ and iterate until the profile has
converged to values where $\rho(r)_n = \rho(r)_{n+1}$.  In Figure
\ref{fig5r}, we show the results of this procedure and find good
agreement with the density and anisotropy profiles measured from the
simulation.  Note that we have not searched the parameter space for
the best match between profiles.  We simply used the same parameters
described above.  A better match would be obtained if we searched
explicitly for the values of $c$, $\beta_E$, and the $q(E)$ power law
parameters which minimize the density profile discrepancy.  More
flexible functions of $q(E)$ could also improve the fit.

That a relatively simple form of the DF matches the results of our
numerical experiments so well justifies the notion that there is
favored outcome through gravitational collapse.  A similar outcome
should apply to all systems formed through radial dissipationless
gravitational collapse, including CDM halos and elliptical galaxies.

\section{Comparison with previous work}
There has been much work attempting to understand the physical nature
of the nearly universal density profiles of CDM halos found in
cosmological N-body simulations.  Next we summarize much of this work
and explain how it relates to the present paper.

One popular way of analytically modeling CDM halo formation is through
so-called secondary infall (SI) models.  This work began with
\citet{1972ApJ...176....1G}, who studied how mass shells accrete onto
a collapsed object.  Later, by imposing self-similar and radial orbits
\citet{1985ApJS...58...39B} and \citet{1984ApJ...281....1F} were able
to derive the inner profile of a halo as a function of the density
profile of the initial perturbation.  More recently, there have been a
number of improvements to these models which include angular momentum
\citep{2010PhRvD..82j4044Z,2010PhRvD..82j4045Z} and generalize to
fully 3D solutions \citep{2011ApJ...734..100L}, including numerical
simulations \citep{2011MNRAS.414.3044V}.  Note that much like our
example in Section 2, the purely radial SI models produce halos which
have steeper central profiles than those observed in cosmological
simulations.  Models which include non-radial velocities are able to
produce NFW or Einasto-like profiles. However, they contain a number
of non-physical assumptions to make the problem tractable, such as
spherical symmetry or assuming that all particles in a given mass
shell have the same velocity.  In this paper, we argue that much of
the physics which determines the density profiles of CDM halos is
unrelated to cosmology.  The details related to cosmological expansion
and accretion included in SI modeling may obscure the important
physics.  Thus, our work compliments the SI approach by identifying
what aspects of the equilibrium state do not depend on cosmology.

Other work focusing on the cosmological context has investigated how
the hierarchical assembly of halos affects the density profile.
\citet{1999ApJ...517...64H} found that a variety of different initial
conditions, including simple monolithic collapse produce similar
density profiles.  \citet{2009MNRAS.396..709W} find that cosmological
hot dark matter (HDM) simulations produce similar profiles to CDM
halos, implying that the mergers due to increased substructure in a CDM
cosmology are not important to structuring density profiles.  This is
also supported by the simulations of \citet{2006ApJ...641..647K}, who
find that the shape of density profiles are preserved during major
mergers of halos in N-body simulations, and earlier work by
\cite{hernquist92,hernquist93,barnes96} who showed that density
profiles similar to those described here arise naturally in
mergers of individual galaxies.
All of this work supports the
idea presented in this paper, that the shape of CDM halos is due to
physical principles common to collapsing systems rather than depending
on the details of initial conditions or cosmology.

It has also been observed in cosmological simulations that the outer
portions of CDM halo density profiles depend on the cosmological
model.  For example the outskirts of halos simulated with vacuum
boundary conditions have profiles steeper than the NFW profile
\citep{1991ApJ...378..496D}.  In the future of a $\Lambda$CDM
universe, halos are steeper than the NFW parameterization at large
radius and are truncated due to the acceleration caused by dark
energy \citep{2005MNRAS.363L..11B}.  We note that the analytic
prediction of $\rho \propto r^{-4}$ breaks down at large radii in this
case since the binding energy of particles is affected by dark energy.
That the outer profile is affected by cosmology supports our claim
that the $\rho \propto r^{-3}$ asymptotic limit in the NFW profile is
a consequence of the cosmological background density and continued
accretion and not due to the primary physics which determines the
profile.

We argue that the ROI is a key physical process which determines the
shape of CDM halo density profiles.  This has been proposed by a
number of other authors and the ROI has been explored both within
N-body simulations \citep{2008ApJ...685..739B,2006ApJ...653...43M} and
semi-analytic models \citep{2005ApJ...634..775B}.  We extend these
studies by proposing a form of the DF motivated by a physical
scenario of gravitational collapse including the ROI.  A different
form of the DF to describe objects formed through gravitational
collapse was put forward by \citet{2005A&A...429..161T} and
\citet{2005A&A...433...57T}.  This DF was
determined by maximizing entropy subject to a constraint involving an
integral of the angular momentum and the energy.  This model generates
density and velocity anisotropy profiles similar to those found in
N-body simulations, but has a feature which does not match in the
detailed form of $f(E,L)$.  Another relevant paper is
\citet{2005ApJ...624L..85M}, where they point out, as we have, that
elliptical galaxy and CDM halo profiles can both be fit by similar
parameterizations (namely the Einasto profile or Sersic model).  In
this paper we have attempted to better explain this connection by
studying a new form of the DF which better matches simulations.

\section{Conclusions}
The similarity between the density profiles of CDM halos and
elliptical galaxies suggests that their formation can be understood in
terms of the same physical processes.  This leads us to propose that
the features of CDM halo density profiles result, to a large extent,
from the physics of gravitational collapse of collisionless systems.
In this paper, the first in a series, we studied these dynamics by
exploring a toy-model, the collapse of a perfectly cold Plummer
sphere.

Using analytic arguments and two numerical experiments, we
demonstrated that the final equilibrium state of our Plummer model
collapse can be understood in fairly simple terms.  In the first
experiment, we ran an N-body simulation of collapse with the particles
constrained to radial orbits.  In the final state, we have found an
inner density profile which is slightly steeper than $\rho \propto
r^{-2}$ and an outer profile which scales as $\rho \propto r^{-4}$.
Simple analytic arguments explain these asymptotic profiles.  In the
second experiment, we relaxed the radial orbit constraint.  Here we
have found the same outer equilibrium profile, but a shallower inner
profile, which we attribute to isotropization introduced by the ROI in
the self-gravitating core of the halo.

We are led to a relatively simple scenario for the formation of
objects through radial gravitational collapse.  Initially
violent-relaxation scatters some particles to unbound orbits and the
remaining weakly bound particles form the outer density profile, which
scales as $r^{-4}$ due to the argument of \citet{1987IAUS..127..511J}
and \citet{1987IAUS..127..339W}.  These particles are not strongly
self-gravitating and thus remain on mostly radial orbits.  The inner
profile is strongly self-gravitating and becomes isotropic through the
onset of ROI resulting in a profile shallower than $\rho \propto
r^{-2}$, like those observed in elliptical galaxies and CDM halos.

We present a form of the DF which matches the final state of our
collapse simulation.  The main features of this model can be related
to our physical picture of collapse; weakly bound (large radii)
particles are on radial orbits and tightly bound (small radii)
particles have isotropic velocities.  This is captured with a term
that describes how circular orbits are as a function of energy,
$h(x,E)$.  These simple considerations are combined with a roughly
Maxwellian energy dependence term, which is a natural expectation
based on statistical arguments.  Our model self-consistently
reproduces the density and anisotropy profiles obtained from our
simulations.  The parameters in the model can also be related to the
bulk properties of the system.

The simple parameterization of the DF motivates a universal form for
density profiles created through this process.  However, non-radial
velocities in the initial conditions of collapse may alter the inner
profile, making it shallower than we find in our numerical experiments.
Continued cosmological infall and accretion could also have some
effect.  These factors may make the inner profiles of CDM halos
sensitive to cosmological parameters
in detail.  Since the outer profile of an
object created through gravitational collapse is predicted to scale as
$\rho \propto r^{-4}$, the shallower profile in CDM halos must be due
to ongoing cosmological accretion and the presence of the cosmological
background.

As evidence for the connection between our toy-model and CDM halos we
find that they both have the same power law for the
pseudo-phase-space density, $\rho/\sigma^3 \propto r^{-1.875}$.  We
also note that at small radii, CDM halos have an anisotropy profile,
$\beta(r)$, which is similar to our collapse model
\citep{2010MNRAS.402...21N}.

The physical picture presented in this paper provides predictions for
CDM halos which can readily be tested.  First, there is no
``universal'' form of the CDM halos in the Universe.  Instead, the
outer halo is altered by accretion and therefore will depend on
background density.  In principle, a halo in a sufficiently
under-dense region will behave as if it is in an open-universe,
causing accretion to halt at earlier epochs allowing the halo to
equilibrate to the $\rho \propto r^{-4}$ form.  Another prediction is
that the final state of halos may have DFs which resemble those from
our collapse simulation.  We will test these predictions in paper 2
(in preparation) and further investigate how the physics of
dissipationless collapse and effects from the cosmological environment
combine to produce the CDM halo density profiles seen in N-body
simulations.

\section{Acknowledgments}
We thank Mark Vogelsberger for useful conversations.  This work was
supported in part by NSF grant AST-0907890 and NASA grants NNX08AL43G
and NNA09DB30A.

\appendix
\section{Distribution Function of the Radial Jaffe Profile}
The distribution function (DF) for a system of purely radial orbits can be
written as
\begin{equation}
f(\vec{x},\vec{v}) = g(\mathcal{E})\delta^D(L^2),
\end{equation}
where $\mathcal{E} \equiv -E$ and $\delta^D$ is a Dirac delta function.  We
derive $g(\mathcal{E})$ for a Jaffe density profile with scale length $a$
and density parameter $\rho_0$,
\begin{equation}
\rho(r) = \frac{\rho_0}{(r/a)^2(1+r/a)^2}.
\end{equation}
We begin with
\begin{equation}
\rho(r) = \int d^3\vec{v} f(\vec{x},\vec{v}) = \int dv_r \frac{g(\mathcal{E})}{r^2},
\end{equation}
where $v_r$ is the radial velocity.  We then perform a change of variables to 
obtain
\begin{equation}
\tilde{\rho}(r)  =\int_0^\Psi \frac{ d\mathcal{E} g(\mathcal{E})}{\sqrt{\Psi - \mathcal{E}}},
\end{equation}
where we have defined $\tilde{\rho}(r) \equiv \sqrt{2} \rho(r) r^2$ and $\Psi
\equiv -\Phi$.  Using the Abel integral equation (see Appendix B in
\citet{2008gady.book.....B}), we can solve for $g(\mathcal{E})$,
\begin{equation}
\label{fintegral}
g(\mathcal{E}) = \frac{1}{\pi}\left [ \int_0^\mathcal{E} \frac{d\Psi}{\sqrt{\mathcal{E} - \Psi}} \frac{d\tilde{\rho}}{d\Psi} + \frac{\tilde{\rho}(\Psi = 0)}{\sqrt{\mathcal{E}}} \right ].
\end{equation}
Since $\Psi$ is a monotonic function of $r$ we are able to use $\tilde{\rho}$ as a function of $\Psi$,
\begin{equation}
\tilde{\rho}(\Psi) = \sqrt{2} \rho_0 a^2 e^{-2 \tilde{\Psi}} \left ( e^{\tilde{\Psi}} - 1  \right )^2,
\end{equation} 
where $\tilde{\Psi} = \Psi /(4\pi\rho_0 a^2 G)$.  Evaluating the
integral in equation (\ref{fintegral}) we obtain,
\begin{equation}
g(\mathcal{E}) = \frac{2a}{\pi^{3/2}}\sqrt{\frac{\rho_0}{G}} \left [\sqrt{2} D \left (\sqrt{\frac{\mathcal{E}}{4\pi a^2\rho_0G}} \right) -  D \left (\sqrt{\frac{\mathcal{E}}{2\pi a^2\rho_0G}} \right) \right ], 
\end{equation}
where $D(x)$ is the Dawson function.


\vskip 0.5 in
\bibliography{paper}
\end{document}